\documentclass[letterpaper]{article} % DO NOT CHANGE THIS
\usepackage[preprint]{aaai2027}  % arXiv preprint: no anonymized-submission notice, real authors shown
\usepackage[hyphens]{url}  % DO NOT CHANGE THIS
\usepackage{graphicx} % DO NOT CHANGE THIS
\usepackage{natbib}  % DO NOT CHANGE THIS AND DO NOT ADD ANY OPTIONS TO IT
\usepackage{caption} % DO NOT CHANGE THIS AND DO NOT ADD ANY OPTIONS TO IT
\usepackage{booktabs} % for better-looking tables
\usepackage{colortbl} % for full-cell result shading
\usepackage{amsmath} % for \text{}, \DeclareMathOperator, equation numbering
\definecolor{tablequality}{HTML}{4472C4}
\definecolor{tablecost}{HTML}{ED7D31}
\definecolor{tableruntime}{HTML}{70AD47}
\newcommand{\qcell}[2]{\cellcolor{tablequality!#1}#2}
\newcommand{\ccell}[2]{\cellcolor{tablecost!#1}#2}
\newcommand{\tcell}[2]{\cellcolor{tableruntime!#1}#2}
\newcommand{\compas}{\texttt{COMPAS}}
\newcommand{\defaultmethod}{\texttt{Default}}
\newcommand{\routellm}{\texttt{RouteLLM}}
\newcommand{\ecotune}{\texttt{EcoTune}}
\newcommand{\promisetune}{\texttt{PromiseTune}}
\newcommand{\gepa}{\texttt{GEPA}}
\newcommand{\smac}{\texttt{SMAC3}}

\newcommand{\alphaevolve}{\texttt{AlphaEvolve}}
\newcommand{\shinkaevolve}{\texttt{ShinkaEvolve}}
\newcommand{\artemis}{\texttt{ARTEMIS}}
\newcommand{\frugalgpt}{\texttt{FrugalGPT}}
\newcommand{\hybridllm}{\texttt{Hybrid-LLM}}
\newcommand{\sweagent}{\texttt{SWE-agent}}
\newcommand{\openhands}{\texttt{OpenHands}}

\newcommand{\ecooptigen}{\texttt{EcoOptiGen}}
\newcommand{\gaivgc}{\texttt{GA4GC}}
\newcommand{\livecodebench}{LiveCodeBench}
\newcommand{\swebench}{SWE-bench}

\DeclareMathOperator*{\argmax}{arg\,max}
\title{\compas{}: Difficulty-Aware Joint Search for Optimizing Code Generation}
\author{
  Jingzhi Gong\textsuperscript{1},
  Jie M. Zhang\textsuperscript{1},
  Gunel Jahangirova\textsuperscript{1},
  Dong Huang\textsuperscript{2},
  Mohammad Reza Mousavi\textsuperscript{1},
  Mark Harman\textsuperscript{3}
}
\affiliations{
  \textsuperscript{1}King's College London, University of London\\
  \textsuperscript{2}National University of Singapore\\
  \textsuperscript{3}University College London, University of London
}

\newboolean{showcomments}
\setboolean{showcomments}{false}
\ifthenelse{\boolean{showcomments}}
  {\newcommand{\nb}[2]{
  \fbox{\bfseries\sffamily\scriptsize#1}
     {\sf\small$\blacktriangleright$\textit{\textcolor{red}{#2}}$\blacktriangleleft$}
   }
  }
  {\newcommand{\nb}[2]{}
   
  }

\begin{document}

\maketitle

\begin{abstract}
Code generation systems make each LLM call with a model, a prompt, and
decoding settings. However, existing optimization methods usually tune only
part of these choices or use one fixed configuration for all tasks: global
optimizers search one configuration for all tasks, routers choose only a
model, and prompt optimizers keep the model and decoding settings fixed. This
leaves their joint, group-specific interactions unclear. We therefore examine
how these choices interact and observe that prompts and decoding settings
interact, tuning effects vary by model, and the best configuration varies by
task difficulty. Guided by these observations, we introduce
\textbf{\compas{}} (\textbf{C}ode-generation \textbf{O}ptimization over
\textbf{M}odels, \textbf{P}rompts, \textbf{A}nd Decoding \textbf{S}ettings), a
difficulty-aware method that learns group-specific quality-cost fronts through
low-cost model selection and joint prompt-decoding search, then routes each
test task to its matching front online without further search. Under a matched
search budget on \livecodebench{}, \compas{} improves pass@1 from 45.9\% for
the best baseline to 52.8\% while reducing cost from \$36.57 to \$4.92. This
also transfers to repository-level code generation on \swebench{}, resolving
76.0\% of tasks versus 70.0\% for the best baseline. Code and the
reproducibility artifact are available at
\url{https://github.com/gjz78910/COMPAS}.
\end{abstract}

\section{Introduction}

\begin{figure*}[t]
\centering
\includegraphics[width=\textwidth]{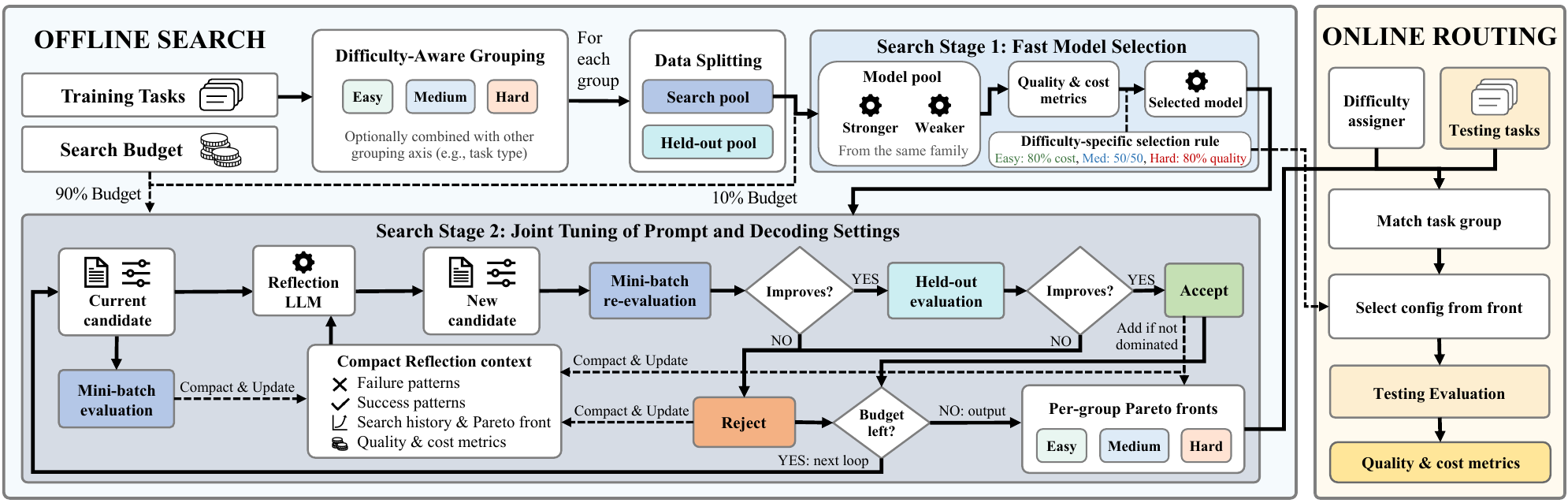}
\caption{\compas{} offline search and online routing workflow.}
\label{fig:overview}
\end{figure*}

Large language models (LLMs) are widely used for code generation, from
function-level synthesis \citep{chen2021humaneval} to repository-scale software
engineering \citep{jimenez2024swebench}. For each LLM call, a code generation
system chooses a model, a prompt, and decoding settings, including temperature,
top-$p$, and frequency penalty \citep{arora2024hyperparams}.

However, code generation workflows usually fix the same choices across
their evaluations \citep{xu2025ecotune,dong26bias}. Because tasks can differ in
difficulty \citep{jain2024livecodebench}, one fixed configuration may
underperform on hard tasks and use more tokens than necessary on easy tasks
\citep{gong25tuning,cheung25comparative}.

Although separate configurations for each task group could address this
trade-off, the search is prohibitively large: 10 prompt candidates, two
models, and 10 values for three decoding settings already produce 60,000
candidates across three groups, requiring over 20 serial days at 30 seconds per evaluation
\citep{xu2025ecotune}. Therefore, a practical method must search this space
efficiently rather than evaluate every candidate.

Existing methods reduce this search in different ways, but each leaves part of
the problem open. \routellm{} \citep{ong2025routellm} selects a model while
keeping the prompt and decoding settings fixed. \gepa{} \citep{agrawal2025gepa}
optimizes prompts but fixes the model and decoding settings, whereas
\ecotune{} \citep{xu2025ecotune} and \ecooptigen{} \citep{wang2023costeffective}
search one global configuration. Similarly, agent systems such as
\sweagent{} \citep{yang2024sweagent} improve a fixed agent harness, while
harness-engineering studies modify that harness \citep{lin2026ahe}.

Consequently, it remains unclear how model, prompt, and decoding settings
interact and whether their joint effects differ by task group. To examine this gap, we conduct a large-scale empirical study across task difficulties, models, prompts, and decoding settings. The study identifies three patterns: prompts and
decoding settings interact, the same tuning update can affect models
differently, and the best observed configuration can differ by task group.

These patterns show that the problem is not merely choosing a stronger model
or prompt, but selecting suitable settings for each group under a limited
budget. We therefore introduce \textbf{\compas{}} (\textbf{C}ode-generation
\textbf{O}ptimization over \textbf{M}odels, \textbf{P}rompts, \textbf{A}nd
Decoding \textbf{S}ettings), a two-phase, difficulty-aware method for LLM and
agent-based code generation. It efficiently selects a model, then jointly searches
prompts and decoding settings offline for each group; at deployment, it routes
each task to a configuration on that group's quality-cost front. Across \livecodebench{} and \swebench{}, \compas{} improves quality by 6.9
and 6.0 percentage points over the strongest baselines, while costing
7.4$\times$ less on \livecodebench{}. In summary, our contributions are:
\begin{enumerate}
\item An empirical characterization of three observations about model,
prompt, and decoding choices for code generation.
\item A difficulty-aware search that partitions problems into groups and
builds a separate quality-cost front per group, rather than a single global
front.
\item A two-stage efficient search that selects the model with a cheap probe,
then uses an LLM-based reflection loop to jointly optimize prompt and
decoding settings.
\end{enumerate}

\section{Empirical Motivations for \compas{}}
\label{sec:prelim}

\begin{figure*}[t]
\centering
\begin{minipage}[t]{0.3\textwidth}
\centering
\includegraphics[width=\textwidth]{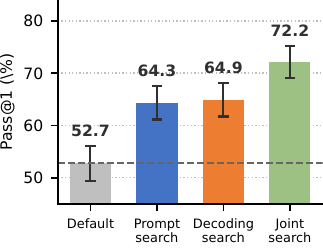}\\
{\small (a) O1: prompts and decoding interact.}
\end{minipage}\hfill
\begin{minipage}[t]{0.3\textwidth}
\centering
\includegraphics[width=\textwidth]{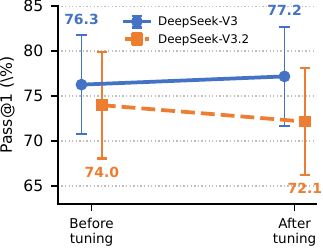}\\
{\small (b) O2: joint tuning is model-specific.}
\end{minipage}\hfill
\begin{minipage}[t]{0.3\textwidth}
\centering
\includegraphics[width=\textwidth]{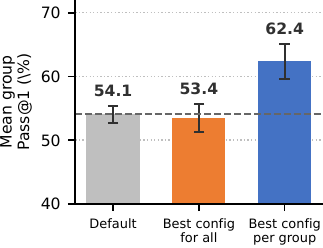}\\
{\small (c) O3: the best config differs by group.}
\end{minipage}
\caption{Three observations from the non-test empirical study that motivate the key designs of \compas{}.}
\label{fig:observations}
\end{figure*}

To examine these interactions without using the primary test tasks, we
evaluate DeepSeek-V3 and DeepSeek-V3.2 on the 880-task release-v5 set with
varied prompts and decoding settings. We report mean Pass@1 and bootstrap
95\% confidence intervals; Figure~\ref{fig:observations} summarizes the
three observations.

\subsection{Observation 1: Prompts and Decoding Interact}
Across all 880 non-test problems and 118{,}965 records, the best recorded
prompt-only and decoding-only choices reach 64.3\% and 64.9\%, whereas the
best joint choice reaches 72.2\% (Figure~\ref{fig:observations}(a)). This
pattern motivates one reflection loop that jointly edits prompts and decoding
settings, avoiding separate reflection calls and candidate evaluations for
the two dimensions.

\subsection{Observation 2: Joint Tuning is Model-Specific}
Prior evidence shows that a prompt optimized for one LLM may fail on another
\citep{gong25tuning}; we therefore test whether the same joint prompt-decoding
update affects models similarly. Figure~\ref{fig:observations}(b) summarizes
876 release-v5 records: the update increases DeepSeek-V3 from 76.3\% to 77.2\%
but lowers DeepSeek-V3.2 from 74.0\% to 72.1\%. This difference motivates selecting a model with a small probe budget before tuning other settings, reducing the subsequent joint search space to one model.

\subsection{Observation 3: The Best Config Differs by Group}
We next split the same 880-problem release into three difficulty groups and
evaluate at least 39 configurations per group. Figure~\ref{fig:observations}(c)
compares the default configuration (54.1\%), the best single configuration
across all groups (53.4\%), and the best configuration per group (62.4\%).
Thus, the best global choice does not improve on the default, whereas
per-group specialization does; this motivates maintaining a separate
quality-cost front per group and routes each test task to its matching front. Together, we build the \compas{} framework around these key design choices.

\section{The \compas{} Framework}
\label{sec:method}

\compas{} follows the offline-optimization and online-deployment pattern used
by LLM optimizers such as \ecotune{} \citep{xu2025ecotune} and model
routers such as \routellm{} \citep{ong2025routellm}: it search offline for a per-group quality-cost front, then, during online
testing, routes each task to a configuration on that front, with no
further search, as shown in Figure~\ref{fig:overview}. 

\subsection{Offline Search Phase}
\label{sec:offline}
Given a set of training tasks and a fixed offline token budget $B$,
\compas{} first partitions the tasks into difficulty groups, then spends
$B$ optimizing each group's model, prompt, and decoding settings in two stages: a fast
model-selection stage followed by joint prompt-decoding tuning.

\subsubsection{Difficulty-Aware Search}
\label{sec:diff-aware}
As \textit{Observation 3} illustrates, task groups prefer different
configurations; hence, a single global configuration $c = (m, t, d) \in \mathcal{M}
\times \mathcal{T} \times \mathcal{D}$, specifying model $m$, prompt $t$,
and decoding settings $d$, cannot be optimal for every
difficulty level. 

\compas{} partitions problems into three difficulty groups $g \in
\mathcal{G} = \{\text{Easy}, \text{Medium}, \text{Hard}\}$ using each
benchmark's own released label: 

\begin{itemize}
    \item \livecodebench{} grades every problem by
Codeforces-style Elo rating (Easy $\le$2000, Medium 2000--3000, Hard
$>$3000) \citep{jain2024livecodebench};
    \item \swebench{} grades every task by an experienced
engineer's estimated fix time (Easy $<$15 min, Medium 15 min--1 hr, Hard
$>$1 hr) \citep{jimenez2024swebench}.
\end{itemize} 

Because both benchmarks provide ordinal difficulty labels, \compas{} reuses
them directly, assigns search-budget weights of 1:2:3 to Easy, Medium,
and Hard such that harder groups receive more search budget, and builds a separate
quality-cost front $F_g$ for each group using its allocated budget.

On \livecodebench{}, \compas{} further separates each difficulty by task
interface because functional and stdin tasks require different prompting and
output contracts \citep{jain2024livecodebench,chen2025selfdebugging}. In our later experiments, this
refined grouping exceeds \texttt{Difficulty-Only} grouping
(52.8\% versus 46.9\%). Thus, on \livecodebench{}, $\mathcal{G}$ ranges over difficulty-type pairs, with $g$ denoting one pair.
Except for this task-interface refinement, \compas{} follows the same search procedure and tunes the same
configuration space on both benchmarks.

\subsubsection{Search Stage 1: Fast Model Selection}
\label{sec:stage1}
In line with \textit{Observation 2}, \compas{} chooses a model for each group before the
expensive joint-tuning stage because the same tuning strategy can affect
models differently.

This stage is skipped when $\mathcal{M}$ has one model. Otherwise,
\compas{} probes every candidate on a random set $S_g$, capped at 8 problems
per group so the shared Stage 1 probe budget cannot eat into Stage 2's
search budget, at a fixed seed configuration $(t_0, d_0)$;
On \livecodebench{}, $S_g$ is also split by problem type, functional or stdin; the probe therefore reports one choice per difficulty-type pair.

The winning model is then fixed by the following cost-aware rule, which
trades off quality against cost using a per-group weight $\alpha_g \in
[0,1]$ (the Online Testing Phase later reuses this same $\alpha_g$ for its own routing rule):
\begin{equation}
m_g^\ast = \argmax_{m \in \mathcal{M}} \ \alpha_g \cdot \widehat{\text{quality}}_m + (1 - \alpha_g) \cdot \widehat{\text{cost}}_m^{-1},
\label{eq:stage1eq}
\end{equation}
where $\widehat{\text{quality}}_m$ and $\widehat{\text{cost}}_m^{-1}$ are
the mean pass rate and inverse cost over $S_g$ at $(t_0,d_0)$, each min-max
normalized across $\mathcal{M}$. Consistent with existing cost-aware routers between
cheaper and costlier models
\citep{chen2023frugalgpt,ding2024hybridllm,ong2025routellm}, we set
$\alpha_g$ to 0.2, 0.5, and 0.8 for Easy, Medium, and Hard. Thus Easy favors
cost, Medium balances both objectives, and Hard favors quality; we
validate this setup against other alternatives in Sensitivity
Analysis.

\subsubsection{Search Stage 2: Joint Tuning}
\label{sec:jointtuning}
Motivated by \textit{Observation 1}, whose strongest measured interaction
is template$\times$decoding, \compas{} jointly mutates the prompt
and decoding settings $d$ on the selected model $m_g^\ast$ in one
reflective loop, via LLM-based reflection, and fits one front per group. Table~\ref{tab:compas-space}
lists the full per-group search space.

Particularly, \compas{} tunes three decoding settings, temperature,
top-$p$, and frequency penalty, following the useful ranges the code
generation study of \citet{arora2024hyperparams} reports, and leaves
presence penalty out of the search space because it penalizes the token
reuse code correctness requires \citep{donato2025studying}.

To search this space without overfitting, \compas{} splits each group's
tasks 80/20 into a search pool and a held-out pool, then spends the
group's search budget evaluating the search pool round-robin rather than
resampling each round, so a tight budget still covers the whole pool.

Each round, \compas{} draws a random mini-batch $M_g$ from the search pool,
mini-batch-evaluates the current candidate $c$ on it, and asks an LLM to
read a compact reflection context (with summarized failure patterns, success patterns,
search history, the current front, and quality-cost metrics). The context
retains at most ten failure and ten success examples, preserving diverse
evidence while bounding the reflection input, before proposing one new
candidate $c'$.

$c'$ is re-evaluated on this same mini-batch $M_g$ using a pass/fail
rule $A_g$: $c'$ passes only if it does not lower score, and either
repairs a previously-failing task or reduces cost within the configured
cost cap, or, for a tied score, repairs a failure, improves a secondary
quality signal, or reduces cost.

\compas{} also draws a random
held-out batch $H_g$ from the held-out pool, and when $H_g$ is nonempty, $c'$ must also pass $A_g$ on $H_g$, preventing it from passing only by overfitting $M_g$. Both $M_g$ and $H_g$ are freshly resampled
each round rather than reused, which avoids the validity loss that
repeatedly querying the same fixed holdout set across many adaptive
rounds would otherwise risk \citep{dwork2015reusable}. Formally,
accept($c'$) iff
\begin{equation}
A_g(M_g,c,c') \land A_g(H_g,c,c').
\label{eq:stage2}
\end{equation}
Subsequently, an accepted $c'$ replaces $c$ and, if not dominated by the
group's front, joins it; otherwise $c$
stays unchanged, and \compas{} repeats the next loop until the group's budget is
spent.

The output of this stage is the group's quality-cost front $F_g$: the
subset of accepted candidates $c = (m_g^\ast, t, d)$ that are
Pareto-optimal on quality and cost, meaning no other accepted candidate
scores at least as well on both.

\begin{table}[!t]
\centering
\small
\begin{tabular}{lp{0.63\columnwidth}}
\toprule
\textbf{Search field} & \textbf{Search range} \\
\midrule
Model & One weaker and one stronger model \\
Prompt & Default and reflection-generated prompts \\
Temperature & $\{0, 0.1, 0.2, 0.3, 0.4, 0.6, 0.8, 1.0\}$ \\
Top-$p$ & $\{0.05, 0.15, 0.35, 0.55, 0.75, 0.95\}$ \\
Frequency penalty & $\{-0.5, -0.1, 0, 0.1, 0.5\}$ \\
\bottomrule
\end{tabular}
\caption{Full \compas{} search space within each task group.}
\label{tab:compas-space}
\end{table}

\subsection{Online Testing Phase}
\label{sec:online}
The group-specific pattern in \textit{Observation 3} motivates \compas{}
to route each test task to its own group's front, rather than using one
fixed configuration for every task. A difficulty assigner first matches
each task $x$ to its difficulty(-type) group $g(x)$ using the existing
label every problem carries \citep{jain2024livecodebench,jimenez2024swebench};
a learned predictor would instead confound our evaluation of \compas{}'s
search-and-routing mechanism with the predictor's own errors; therefore, the main results use official labels to isolate that mechanism \citep{ong2025routellm}. The Sensitivity Analysis tests this assumption directly with a learned
\texttt{XGBoost-Router}: routing by predicted rather than official labels
reaches 49.7\% Pass@1, only 3.1 points below the label-based 52.8\% (not significant, paired Wilcoxon $p=0.4732$), suggesting that
deployment need not depend on released labels.

\compas{} then selects, from $g(x)$'s front $F_{g(x)}$, the configuration
maximizing a normalized quality-cost trade-off, reusing the same
$\alpha_g$ weights fixed in Stage 1 (Eq.~\ref{eq:stage1eq}):
\begin{equation}
\rho(x) = c^\ast(x) = \argmax_{c \in F_{g(x)}} \ \alpha_{g(x)} \cdot \widehat{\text{quality}}(c) + (1 - \alpha_{g(x)}) \cdot \widehat{\text{cost}}(c)^{-1},
\label{eq:route}
\end{equation}
where quality and inverse cost are min-max normalized over $F_{g(x)}$; if
$F_{g(x)}$ is empty, \compas{} falls back to a coarser difficulty front,
then the fixed default. This is a lookup over an already-built front, so
no search happens online.

\section{Experimental Setup}
\label{sec:setup}

\subsection{Model Setup}
We evaluate three model families: DeepSeek-V4-Flash and DeepSeek-V4-Pro
\citep{deepseek2026v4}, Devstral Small 2 and Devstral 2
\citep{mistral2025devstral2}, and Qwen3.5-Flash and Qwen3.5-Plus
\citep{alibaba2026qwen35}. Each pool holds one weaker and one stronger
model from the same family, since within-family capability scales
predictably \citep{kaplan2020scaling}, letting the Stage 1 search
reliably rank them without the added confound of cross-provider
differences.
We use LLM-based reflection
\citep{agrawal2025gepa} to propose new joint prompt-decoding candidates,
and fix DeepSeek-V4-Pro as the reflection model due to its strong
reasoning capability at moderate API cost \citep{deepseek2026v4}.

\begin{table}[!t]
\centering
\small
\begin{tabular}{lll}
\toprule
\textbf{Benchmark} & \textbf{Train set} & \textbf{Test set} \\
\midrule
\livecodebench{} & Release v5 (880) & v6-only (175) \\
\livecodebench{} & Random (90) & Random (90) \\
\swebench{} Verified & Random (44) & Verified-mini (50) \\
\bottomrule
\end{tabular}
\caption{Benchmark task sets; 44 \swebench{} tasks
are deduplicated from a 50-task random split.}
\label{tab:datasets}
\end{table}

\subsection{\livecodebench{} Setup}
As summarized in Table~\ref{tab:datasets}, we evaluate on \livecodebench{} \citep{jain2024livecodebench}, whose numbered releases add problems collected after each prior release's cutoff specifically to guard against training-data contamination, over both task
types (functional and standard-input): we train on release v5 and test on
the 175 problems added in release v6 (v6-only), which by construction
postdate every v5 training problem, consistent with recent studies \citep{sharifloo25where, jia2025specfix, xu2026trinity, chai2026gxpo}. The second split is an independent random split with 90
training and 90 test tasks \citep{cosplay2026, white2025livebench}. During
search, we generate one code output per problem, whereas the final
evaluation generates ten code outputs per problem; at temperature zero, one deterministic
output is generated and counted ten times. We report the mean single-output pass rate
\begin{equation}
\operatorname{Pass@1} = \frac{1}{|\mathcal{P}|}\sum_{i\in\mathcal{P}}
\frac{1}{n_i}\sum_{j=1}^{n_i} z_{ij},
\end{equation}
where $n_i$ is the number of outputs evaluated for problem $i$ and $z_{ij}=1$
when output $j$ passes its benchmark tests, following \livecodebench{}
\citep{jain2024livecodebench}.
Notably, our setup supports a controlled method comparison, but not a
leaderboard claim under our study-specific protocol.

\subsection{\swebench{} Setup}
\label{sec:swebench-setup}
We evaluate on \swebench{} Verified-mini, the curated 50-task subset of
\swebench{} Verified, because it supports a controlled repository-level
evaluation at practical cost \citep{jiang25issue,tripathy2026swenergy,zhang2026dgm,kapoor2026holistic,li2025swedeba}, as shown in Table~\ref{tab:datasets}. To keep search disjoint from testing, we randomly sample
50 Verified tasks and remove the overlap tasks, leaving a 44-task search pool with 20 Easy, 16 Medium, and 8
Hard tasks. 

We run all agents in official \swebench{} Docker images using the fixed
\texttt{mini-SWE-agent} harness, whose lightweight Bash loop is used in
controlled \swebench{} studies to separate model behavior from scaffold choices
\citep{chen2026agenttests,tripathy2026swenergy}. Additionally, we generate
one patch per task, following the pass@1 protocol used by
\citet{yang2024sweagent} and \citet{zhang2026dgm}.
DeepSeek~V4~series models use no-thinking mode because their thinking mode ignores decoding
settings \citep{deepseek2026v4}.

\subsection{Baselines}
We choose recent leading methods from
complementary optimization categories: (1) \textbf{\defaultmethod{}} fixes the stronger LLM as the
    no-search reference, following the default model, prompt, and decoding settings for both benchmarks \citep{jain2024livecodebench,yang2024sweagent}; (2) \textbf{\routellm{}} learns a cost-quality router from preference
    data to select a strong or weak model for each query \citep{ong2025routellm}; (3) \textbf{\ecotune{}} tunes decoding hyperparameters using a dynamic fidelity schedule that adapts evaluation depth to the
    remaining token budget \citep{xu2025ecotune}; and (4) \textbf{\promisetune{}} combines causal analysis with explainable
    rules to identify causally promising configuration settings
    \citep{chen2026promisetune}.

% \begin{itemize}
%     \item \textbf{\defaultmethod{}} fixes the stronger LLM as the
%     no-search reference, following the default model, prompt, and decoding settings for both benchmarks \citep{jain2024livecodebench,yang2024sweagent}.
%     \item \textbf{\routellm{}} learns a cost-quality router from preference
%     data to select a strong or weak model for each query \citep{ong2025routellm}.
%     \item \textbf{\ecotune{}} tunes decoding hyperparameters using a dynamic fidelity schedule that adapts evaluation depth to the
%     remaining token budget \citep{xu2025ecotune}.
%     \item \textbf{\promisetune{}} combines causal analysis with explainable
%     rules to identify causally promising configuration settings
%     \citep{chen2026promisetune}.
% \end{itemize}

The three search baselines are state-of-the-art in related domains, reporting gains or advantages
over widely used methods such as \texttt{FrugalGPT} \citep{chen2023frugalgpt} and
\texttt{Hybrid-LLM} \citep{ding2024hybridllm} for routing, \texttt{BOHB}
\citep{falkner2018bohb} for HPO, and \texttt{SMAC3} \citep{lindauer2022smac3} for
configuration tuning. 

To keep this comparison fair, all methods use the same evaluation protocol and the
same search token-budget cap: 1M tokens on \livecodebench{} and 4M tokens on
\swebench{}, calibrated from preliminary experiment traces. This cap
bounds only the offline search phase; reported dollar totals also
include uncapped online deployment, and differ across methods mainly
because each routes to models with different per-token prices, not
because of unequal search budgets. For fairness,
\ecotune{} and \promisetune{} search the same model, prompt, and decoding
dimensions as \compas{}, whereas \routellm{} keeps prompts and decoding fixed
by design.

For each single-seed experiment, we use two-sided paired Wilcoxon signed-rank
tests, Vargha--Delaney $A_{12}$ effect sizes, and paired-bootstrap 95\%
confidence intervals on matched per-problem outcomes to assess paired differences and report their
effect size and uncertainty \citep{dror2018hitchhiker}. Scott--Knott Effect Size Difference (ESD) ranks
are used only for the five independent v6-only seed aggregates because they
require independent runs
\citep{tantithamthavorn2018impact,scottknottesd}.

\section{Experiments}
\label{sec:results}

\begin{table*}[!t]
\centering
\small
\setlength{\tabcolsep}{1mm}
\begin{tabular}{@{}lrrr||rrr||rrrrrr@{}}
\toprule
& \multicolumn{3}{c}{\textbf{v6-only (main run)}}
& \multicolumn{3}{c}{\textbf{Random split}}
& \multicolumn{6}{c}{\textbf{v6-only (five seeds)}} \\
\cmidrule(lr){2-4}\cmidrule(lr){5-7}\cmidrule(lr){8-13}
\textbf{Method} & \textbf{Pass@1 (\%)} & \textbf{Cost} & \textbf{Time}
& \textbf{Pass@1 (\%)} & \textbf{Cost} & \textbf{Time}
& \textbf{Pass@1 (\%)} & \textbf{$r_p$} & \textbf{Cost} & \textbf{$r_c$} & \textbf{Time} & \textbf{$r_t$} \\
\midrule
\defaultmethod{}
& \qcell{5.0}{40.6 [33.1, 48.0]} & \ccell{17.6}{\$14.67} & \tcell{25.0}{\textbf{6.9h}}
& \qcell{13.0}{55.6 [45.6, 65.6]} & \ccell{19.7}{\$5.04} & \tcell{25.0}{\textbf{2.7h}}
& \qcell{6.1}{41.4 $\pm$ 1.2} & \qcell{6.1}{4} & \ccell{16.8}{\$71.82} & \ccell{16.8}{3} & \tcell{25.0}{\textbf{37.6h}} & \tcell{25.0}{\textbf{1}} \\
\routellm{}
& \qcell{10.3}{42.9 [35.4, 50.3]} & \ccell{23.2}{\$7.78} & \tcell{20.6}{38.2h}
& \qcell{13.0}{55.6 [44.4, 65.6]} & \ccell{25.0}{{\$2.19}} & \tcell{24.0}{5.4h}
& \qcell{12.9}{44.8 $\pm$ 1.7} & \qcell{12.9}{2} & \ccell{21.4}{\$39.09} & \ccell{21.4}{2} & \tcell{19.4}{192.0h} & \tcell{19.4}{3} \\
\ecotune{}
& \qcell{17.4}{45.9 [39.2, 52.6]} & \ccell{5.0}{\$36.57} & \tcell{5.0}{149.1h}
& \qcell{5.0}{51.1 [41.1, 61.1]} & \ccell{7.0}{\$11.86} & \tcell{23.0}{8.2h}
& \qcell{5.0}{40.8 $\pm$ 4.6} & \qcell{5.0}{5} & \ccell{5.0}{\$161.82} & \ccell{5.0}{5} & \tcell{5.0}{586.8h} & \tcell{5.0}{5} \\
\promisetune{}
& \qcell{7.7}{41.7 [34.3, 49.1]} & \ccell{17.2}{\$15.40} & \tcell{24.4}{11.4h}
& \qcell{20.2}{59.6 [50.0, 68.7]} & \ccell{5.0}{\$12.94} & \tcell{5.0}{57.4h}
& \qcell{6.9}{41.8 $\pm$ 2.4} & \qcell{6.9}{3} & \ccell{12.2}{\$108.13} & \ccell{12.2}{4} & \tcell{10.2}{442.9h} & \tcell{10.2}{4} \\
\textbf{\compas{}}
& \qcell{25.0}{\textbf{52.8 [45.6, 60.2]}} & \ccell{25.0}{\textbf{\$4.92}} & \tcell{24.6}{9.9h}
& \qcell{25.0}{\textbf{62.2 [52.2, 72.2]}} & \ccell{25.0}{\textbf{\$1.09}} & \tcell{24.0}{5.5h}
& \qcell{25.0}{\textbf{51.6 $\pm$ 1.6}} & \qcell{25.0}{\textbf{1}} & \ccell{25.0}{\textbf{\$11.84}} & \ccell{25.0}{\textbf{1}} & \tcell{24.6}{48.9h} & \tcell{24.6}{2} \\
\bottomrule
\end{tabular}
\caption{Main effectiveness and robustness on \livecodebench{}. Main-run
Pass@1 values are means [bootstrap 95\% CI], while five-seed Pass@1 values are
mean $\pm$ Standard Deviation. Cost and Time are total API cost and actual LLM
wall-clock time, summed over all runs. $r$ is
Scott--Knott ESD rank over independent seeds. Darker color and bold indicate better
performance.}
\label{tab:lcb-main}
\end{table*}

\subsection{Effectiveness on \livecodebench{}}
\label{sec:main-effectiveness}

\subsubsection{Main Effectiveness}
We first ask the central question: under a matched search budget, every
method capped at the same number of search tokens, does \compas{}
improve the quality-cost tradeoff on \livecodebench{}? Table~\ref{tab:lcb-main}
(left) reports Pass@1, cost, and time on the main v6-only split
\citep{jia2025specfix, xu2026trinity, chai2026gxpo}; because the cap is
in tokens, not hours, wall-clock time still varies widely across methods
depending on how quickly each spends that shared budget.

Overall, \compas{} attains the
highest Pass@1 (52.8\%) and the lowest cost (\$4.92) of all five methods; while the strongest Pass@1 baseline, \ecotune{}, reaches 45.9\% with \$36.57 (paired
Wilcoxon, $p{=}0.0167$; $A_{12}{=}0.544$).

\emph{\textbf{Takeaway.} \compas{} attains the strongest observed quality-cost trade-off on the primary split.}

\subsubsection{Robustness to Dataset Randomness}
The main run leaves open whether the gain is specific to the v6-only
split of problems. To check robustness, we repeat the same five-method
comparison on an independent random split of the same benchmark
\citep{cosplay2026, white2025livebench}.

As shown in Table~\ref{tab:lcb-main} (middle), \compas{} again leads on
Pass@1 (62.2\%), ahead of the second-best method \promisetune{} (59.6\%), and does so at 91.5\% lower cost (\$1.09 versus
\$12.94). It also costs less than \routellm{} (\$2.19) while achieving
11.9\% relatively higher Pass@1 (62.2\% versus 55.6\%).

\emph{\textbf{Takeaway.} On the independent random split, \compas{} again attains the highest
Pass@1 at the lowest cost.}

\begin{table*}[!t]
\centering
\small
\setlength{\tabcolsep}{1mm}
\begin{tabular}{lrc|rc||rc|rc||rc|rc}
\toprule
& \multicolumn{4}{c}{\textbf{Easy}} & \multicolumn{4}{c}{\textbf{Medium}} & \multicolumn{4}{c}{\textbf{Hard}} \\
\cmidrule(lr){2-5}\cmidrule(lr){6-9}\cmidrule(lr){10-13}
& \multicolumn{2}{c}{\textbf{Functional ($n{=}17$)}} & \multicolumn{2}{c}{\textbf{Stdin ($n{=}26$)}}
& \multicolumn{2}{c}{\textbf{Functional ($n{=}26$)}} & \multicolumn{2}{c}{\textbf{Stdin ($n{=}26$)}}
& \multicolumn{2}{c}{\textbf{Functional ($n{=}20$)}} & \multicolumn{2}{c}{\textbf{Stdin ($n{=}60$)}} \\
\textbf{Method} & \textbf{Pass@1 (\%)} & \textbf{$r$} & \textbf{Pass@1 (\%)} & \textbf{$r$} & \textbf{Pass@1 (\%)} & \textbf{$r$} & \textbf{Pass@1 (\%)} & \textbf{$r$} & \textbf{Pass@1 (\%)} & \textbf{$r$} & \textbf{Pass@1 (\%)} & \textbf{$r$} \\
\midrule
\defaultmethod{} & \qcell{6}{28.2$\pm$4.9} & \qcell{6}{4} & \qcell{5}{98.5$\pm$2.1} & \qcell{5}{4} & \qcell{8}{8.5$\pm$1.7} & \qcell{8}{2} & \qcell{21}{75.4$\pm$4.4} & \qcell{21}{2} & \qcell{16}{7.0$\pm$4.5} & \qcell{16}{3} & \qcell{5}{31.3$\pm$1.4} & \qcell{5}{4} \\
\routellm{} & \qcell{6}{30.6$\pm$2.6} & \qcell{6}{2} & \qcell{5}{98.5$\pm$2.1} & \qcell{5}{4} & \qcell{7}{7.7$\pm$3.8} & \qcell{7}{3} & \qcell{25}{\textbf{76.9$\pm$5.4}} & \qcell{25}{\textbf{1}} & \qcell{16}{7.0$\pm$2.7} & \qcell{16}{3} & \qcell{25}{\textbf{40.3$\pm$2.5}} & \qcell{25}{\textbf{1}} \\
\ecotune{} & \qcell{5}{26.2$\pm$2.1} & \qcell{5}{5} & \qcell{17}{99.4$\pm$0.6} & \qcell{17}{2} & \qcell{6}{7.2$\pm$1.1} & \qcell{6}{4} & \qcell{5}{70.1$\pm$8.9} & \qcell{5}{5} & \qcell{5}{4.4$\pm$2.9} & \qcell{5}{4} & \qcell{10}{33.7$\pm$8.2} & \qcell{10}{3} \\
\promisetune{} & \qcell{6}{29.3$\pm$0.9} & \qcell{6}{3} & \qcell{11}{98.9$\pm$1.6} & \qcell{11}{3} & \qcell{5}{6.8$\pm$1.7} & \qcell{5}{5} & \qcell{12}{72.4$\pm$5.9} & \qcell{12}{4} & \qcell{18}{7.5$\pm$1.8} & \qcell{18}{2} & \qcell{11}{33.9$\pm$5.1} & \qcell{11}{3} \\
\textbf{\compas{}} & \qcell{25}{\textbf{97.6$\pm$3.2}} & \qcell{25}{\textbf{1}} & \qcell{25}{\textbf{100.0$\pm$0.0}} & \qcell{25}{\textbf{1}} & \qcell{25}{\textbf{16.5$\pm$8.7}} & \qcell{25}{\textbf{1}} & \qcell{15}{73.6$\pm$4.0} & \qcell{15}{3} & \qcell{25}{\textbf{9.0$\pm$4.2}} & \qcell{25}{\textbf{1}} & \qcell{18}{37.3$\pm$2.5} & \qcell{18}{2} \\
\bottomrule
\end{tabular}
\caption{Per-group Pass@1 (\%, mean $\pm$ SD over the same five seeds as
Table~\ref{tab:lcb-main}) on \livecodebench{} v6-only. $r$ column is Scott--Knott ESD rank; formatting follows Table~\ref{tab:lcb-main}.}
\label{tab:group-breakdown}
\end{table*}

\subsubsection{Robustness to Internal Randomness}
The random-split check controls for which tasks are tested, but not for
randomness in \compas{}'s own offline search: a different search seed
could still reshape the offline-built fronts. We therefore repeat the
full v6-only
comparison on four additional seeds; Table~\ref{tab:lcb-main} (right)
reports the five-seed aggregate, ranked with
Scott--Knott ESD test. {In particular, \compas{} is the only rank 1 for both Pass@1 and cost
across all five seeds, with a pass@1 win over the best baseline
(\routellm{}, 44.8\%) of 15.2\% and a cost saving of 69.7\% (\$11.84
versus \$39.09).}

Looking into the six difficulty-type groups in
Table~\ref{tab:group-breakdown}, \compas{} leads most clearly on functional
tasks: it reaches 97.6\% on Easy/functional and 16.5\% on
Medium/functional, versus at most 30.6\% and 8.5\% for any baseline,
respectively. On Hard/functional, its raw Pass@1 is also higher
(9.0\% versus 7.5\%) and it receives the sole best Scott--Knott ESD rank
across the five independent search seeds. By contrast, on Medium/stdin and
Hard/stdin, \routellm{}'s per-difficulty model routing performs better;
however, \compas{} remains much cheaper overall
(\$11.84 versus \$39.09; Table~\ref{tab:lcb-main}, right).

\emph{\textbf{Takeaway.} The main-run advantage holds
across the randomness in its own search.}

\begin{table*}[!t]
\centering
\small
\setlength{\tabcolsep}{1mm}
\begin{tabular}{@{}lrrrrrr||rrr||rrr@{}}
\toprule
& \multicolumn{6}{c}{\textbf{\swebench{} Verified-mini}}
& \multicolumn{3}{c}{\textbf{Devstral}}
& \multicolumn{3}{c}{\textbf{Qwen3.5}} \\
\cmidrule(lr){2-7}\cmidrule(lr){8-10}\cmidrule(lr){11-13}
\textbf{Method} & \textbf{Resolve (\%)} & \textbf{Easy} & \textbf{Med} & \textbf{Hard} & \textbf{Cost} & \textbf{Time}
& \textbf{Pass@1 (\%)} & \textbf{Cost} & \textbf{Time} & \textbf{Pass@1 (\%)} & \textbf{Cost} & \textbf{Time} \\
\midrule
\defaultmethod{}
& \qcell{11.7}{68.0 (34/50)} & \qcell{5.0}{16/19} & \qcell{17.0}{17/23} & \qcell{5.0}{1/8} & \ccell{9.4}{\$31.09} & \tcell{6.4}{10.3h}
& \qcell{5.0}{20.6 [14.9, 26.9]} & \ccell{25.0}{\textbf{\$0.11}} & \tcell{25.0}{\textbf{0.1h}}
& \qcell{14.9}{33.1 [26.3, 40.0]} & \ccell{19.5}{\$6.29} & \tcell{25.0}{\textbf{3.3h}} \\
\routellm{}
& \qcell{15.0}{70.0 (35/50)} & \qcell{5.0}{16/19} & \qcell{17.0}{17/23} & \qcell{25.0}{\textbf{2/8}} & \ccell{25.0}{\textbf{\$8.82}} & \tcell{5.0}{10.6h}
& \qcell{5.0}{20.6 [14.9, 26.3]} & \ccell{13.7}{\$2.21} & \tcell{22.5}{1.5h}
& \qcell{5.0}{25.7 [19.4, 32.0]} & \ccell{22.4}{\$3.52} & \tcell{24.7}{4.2h} \\
\ecotune{}
& \qcell{5.0}{64.0 (32/50)} & \qcell{25.0}{\textbf{17/19}} & \qcell{5.0}{14/23} & \qcell{5.0}{1/8} & \ccell{10.3}{\$29.80} & \tcell{21.4}{7.0h}
& \qcell{25.0}{\textbf{29.3 [23.0, 35.7]}} & \ccell{5.0}{\$3.82} & \tcell{5.0}{11.1h}
& \qcell{17.5}{35.0 [28.5, 41.7]} & \ccell{5.0}{\$20.25} & \tcell{5.0}{59.8h} \\
\promisetune{}
& \qcell{15.0}{70.0 (35/50)} & \qcell{5.0}{16/19} & \qcell{17.0}{17/23} & \qcell{25.0}{\textbf{2/8}} & \ccell{13.0}{\$25.90} & \tcell{25.0}{\textbf{6.2h}}
& \qcell{17.2}{25.9 [19.9, 31.9]} & \ccell{6.5}{\$3.54} & \tcell{11.2}{7.7h}
& \qcell{6.6}{26.9 [20.6, 33.7]} & \ccell{19.7}{\$7.08} & \tcell{23.8}{6.7h} \\
\textbf{\compas{}}
& \qcell{25.0}{\textbf{76.0 (38/50)}} & \qcell{25.0}{\textbf{17/19}} & \qcell{25.0}{\textbf{19/23}} & \qcell{25.0}{\textbf{2/8}} & \ccell{5.0}{\$37.37} & \tcell{22.3}{6.8h}
& \qcell{19.9}{27.1 [20.7, 33.5]} & \ccell{18.2}{\$1.38} & \tcell{11.9}{7.3h}
& \qcell{25.0}{\textbf{40.6 [33.9, 47.4]}} & \ccell{25.0}{\textbf{\$1.01}} & \tcell{20.7}{15.3h} \\
\bottomrule
\end{tabular}
\caption{Generalization across repo-level benchmark and model family. \swebench{} reports
resolve rate and solved counts; Devstral and Qwen3.5 report mean Pass@1
[bootstrap 95\% CI]. Formatting follows Table~\ref{tab:lcb-main}.}
\label{tab:generalization}
\end{table*}

\subsection{Generalization}
\label{sec:generalization}
We next evaluate whether the gain holds outside its
primary setting, across repository-level tasks (agent-based code generation on \swebench{}) and model family (DeepSeek to Devstral and
Qwen3.5). Table~\ref{tab:generalization} reports both axes.

\subsubsection{Generalization to Repository-Level Tasks}
On \swebench{} Verified-mini
\citep{li2025swedeba, jiang25issue, kapoor2026holistic} (Table~\ref{tab:generalization}, left), {\compas{}
resolves 38/50 tasks (76.0\%), ahead of the 70.0\% achieved by
\routellm{} and \promisetune{}, but at a higher cost (\$37.37).} Because
Verified-mini is a curated subset, these numbers are not comparable to the full leaderboard
results.

The per-group columns show where this advantage comes from: \compas{} resolves 19/23 Medium tasks, versus 17/23 for \defaultmethod{}, \routellm{}, and \promisetune{} and 14/23 for \ecotune{}; it also resolves 17/19 Easy tasks, matching \ecotune{} and exceeding the other baselines, while Hard remains low for all methods (at most 2/8). Moreover, by tracing the logs, the
higher price is because \compas{} routes more tasks to the higher-capability, pricier models.

\emph{\textbf{Takeaway.} \compas{} resolves the most \swebench{} tasks, driven by stronger Easy and Medium results.}

\subsubsection{Generalization Across Model Family}
On Qwen3.5 (Table~\ref{tab:generalization}, right), \compas{} attains the
highest Pass@1 (40.6\%) and lowest cost (\$1.01). On Devstral
(Table~\ref{tab:generalization}, middle), it improves on \defaultmethod{}
(27.1\% versus 20.6\%) but remains below \ecotune{} (29.3\%), the only
setting where \compas{} does not lead. 

The Devstral gap traces to \compas{}'s cheap model probe: Devstral's two
pool models score too close together for the probe to tell apart, so
the cost-aware rule breaks the tie toward the cheaper one without
confirming it truly matches the larger model; on Qwen3.5 and DeepSeek-V4,
the pool models are easier to distinguish on the same cheap probe, so
this failure mode does not arise. 

\emph{\textbf{Takeaway.} Across model families, \compas{} leads on Qwen3.5 and improves on the default on Devstral.}

\subsection{Component Ablation}
\label{sec:ablation}
Table~\ref{tab:ablation} compares full \compas{} against six variants,
each removing one component, to measure each component's contribution
under the shared budget and evaluation protocol.
\texttt{No-Prompt-Search} and \texttt{No-Decoding-Search} each disable one half of the
joint search, isolating whether prompts and
decoding settings must be searched together. \texttt{No-History-Pack} disables
the persistent record of prior rounds' failures, so reflection uses feedback
only from the current round. \texttt{No-History-Compacting} retains this
history without its compaction operation. \texttt{No-Model-Selection} disables the cheap model probe and fixes DeepSeek-V4-Flash;
\texttt{No-Task-Grouping} uses one configuration for all difficulty levels.

\begin{table}[!t]
\centering
\small
\setlength{\tabcolsep}{1mm}
\begin{tabular}{lrrr}
\toprule
\textbf{Variant} & \textbf{Pass@1 (\%)} & \textbf{Cost} & \textbf{Time} \\
\midrule
\texttt{No-Decoding-Search} & \qcell{15.3}{46.3 [38.9, 53.7]} & \ccell{5.9}{\$15.33} & \tcell{8.5}{11.4h} \\
\texttt{No-Prompt-Search} & \qcell{8.4}{41.7 [34.3, 49.1]} & \ccell{5.8}{\$15.36} & \tcell{5.0}{12.4h} \\
\texttt{No-History-Pack} & \qcell{7.6}{41.1 [33.7, 48.6]} & \ccell{6.4}{\$15.01} & \tcell{5.8}{12.2h} \\
\texttt{No-History-Compacting} & \qcell{21.2}{50.3 [42.9, 57.7]} & \ccell{24.8}{\$4.81} & \tcell{21.9}{7.3h} \\
\texttt{No-Task-Grouping} & \qcell{5.0}{39.4 [32.6, 46.9]} & \ccell{5.0}{\$15.81} & \tcell{5.4}{12.3h} \\
\texttt{No-Model-Selection} & \qcell{11.8}{44.0 [36.6, 51.4]} & \ccell{25.0}{\textbf{\$4.71}} & \tcell{25.0}{\textbf{6.3h}} \\
\textbf{Full \compas{}} & \qcell{25.0}{\textbf{52.8 [45.6, 60.2]}} & \ccell{24.6}{\$4.92} & \tcell{13.4}{9.9h} \\
\bottomrule
\end{tabular}
\caption{Ablation study on LCB v6-only; each row disables
the named \compas{} operation; formatting follows Table~\ref{tab:lcb-main}.}
\label{tab:ablation}
\end{table}

{Overall, every variant scores below full \compas{} (52.8\%).} Removing
decoding or prompt search lowers Pass@1 to 46.3\% and 41.7\%,
so neither dimension alone is sufficient. Without the history pack, Pass@1
falls to 41.1\%;
likewise, fixing one model before search yields 44.0\%, supporting model
probing. Finally, \texttt{No-Task-Grouping} hurts the most at 39.4\%, below
\defaultmethod{}: one configuration loses the specialization provided by
per-group fronts.

In contrast, several ablations cost \$15.01--\$15.81 rather than the
\$4.71--\$4.92 of the other variants, reflecting differences in model selection across variants; so we do not draw any cost conclusions here.

\emph{\textbf{Takeaway.} Each component contributes to the reported full \compas{} results; task grouping has the largest reduction.}

\subsection{Sensitivity Analysis}
\label{sec:sensitivity}
Table~\ref{tab:robustness} varies five design axes around
\compas{}'s default implementation; unlike the components tested in the
ablation study above, these are decisions that a deployer sets before running
\compas{} on a new setting. Grouping tests \texttt{Difficulty-Only},
collapsing the default difficulty$\times$type split into difficulty
alone. Predictor
tests two difficulty assigners for the ground-truth difficulty
labels: \texttt{XGBoost-Router}, a
gradient-boosted classifier trained on problem embeddings, and
\texttt{KNN-Router}, a majority vote of difficulty labels among the $k$
nearest training problems. Budget tests \texttt{Half-Budget} (0.5M
tokens) and \texttt{Double-Budget} (2M tokens).
Reflection tests \texttt{Short-Context} (2/2 failure/success examples
instead of 10/10) and \texttt{Flash-Reflect} (a cheaper reflection
model in place of the default). Pareto selection tests \texttt{Max-Pass}
and \texttt{Min-Cost}, choosing a front's best-quality or cheapest point
instead of the default quality-cost rule.

\begin{table}[!t]
\centering
\small
\setlength{\tabcolsep}{1mm}
\begin{tabular}{llrrr}
\toprule
\textbf{Axis} & \textbf{Variant} & \textbf{Pass@1 (\%)} & \textbf{Cost} & \textbf{Time} \\
\midrule
Grouping & \texttt{Difficulty-Only} & \qcell{17.9}{46.9 [39.4, 54.3]} & \ccell{25.0}{\textbf{\$1.69}} & \tcell{22.0}{9.8h} \\
Predictor & \texttt{XGBoost-Router} & \qcell{25.0}{49.7 [42.3, 57.1]} & \ccell{20.4}{\$4.73} & \tcell{25.0}{{8.2h}} \\
Predictor & \texttt{KNN-Router} & \qcell{13.6}{45.1 [37.7, 52.6]} & \ccell{25.0}{\$1.72} & \tcell{23.2}{9.2h} \\
Budget & \texttt{Half-Budget} & \qcell{5.0}{41.7 [34.8, 49.1]} & \ccell{5.5}{\$14.64} & \tcell{20.7}{10.5h} \\
Budget & \texttt{Double-Budget} & \qcell{17.3}{46.6 [39.3, 54.0]} & \ccell{5.0}{\$14.95} & \tcell{5.0}{18.8h} \\
Reflection & \texttt{Short-Context} & \qcell{9.0}{44.0 [36.6, 51.4]} & \ccell{24.9}{\$1.73} & \tcell{21.6}{9.3h} \\
Reflection & \texttt{Flash-Reflect} & \qcell{20.5}{50.3 [42.9, 57.7]} & \ccell{20.5}{\$4.64} & \tcell{25.0}{\textbf{6.7h}} \\
Pareto & \texttt{Max-Pass} & \qcell{21.5}{50.9 [43.4, 58.3]} & \ccell{20.6}{\$4.61} & \tcell{18.1}{10.9h} \\
Pareto & \texttt{Min-Cost} & \qcell{20.5}{50.3 [42.9, 57.7]} & \ccell{20.5}{\$4.68} & \tcell{18.1}{10.9h} \\
Default & \textbf{\compas{}} & \qcell{25.0}{\textbf{52.8 [45.6, 60.2]}} & \ccell{20.1}{\$4.92} & \tcell{19.8}{9.9h} \\
\bottomrule
\end{tabular}
\caption{Sensitivity analysis on LCB v6-only around key \compas{} design choices; formatting follows
Table~\ref{tab:lcb-main}.}
\label{tab:robustness}
\end{table}

Confirming that the extra granularity is worth it, \texttt{Difficulty-Only}
grouping scores lower (46.9\% versus 52.8\%) because it collapses the
functional/stdin split that the largest per-group gains above are traced
to into one group with a single, less specialized configuration.

Among the budget settings, 1M tokens scores highest (52.8\%), while
0.5M scores lower (41.7\%) and doubling the budget to 2M
does not significantly improve on 1M (46.6\%; confidence
intervals overlap). Given this and the scale of our experiments,
we fix 1M tokens as \compas{}'s default offline budget. The 2M run's Pareto fronts
explain what happened here: two groups score only 31.7\% and 18.0\% Pass@1, overfitting their search samples, so in
this run, more search only gave the accept/reject gate more chances to
commit to an overfit candidate rather than a better one.

On the remaining axes, the 3.1-point \texttt{XGBoost-Router} gap against the main result is not significant ($p=0.4732$), suggesting a learned difficulty predictor could approximate the difficulty labels in real deployment, unlike the 7.7-point \texttt{KNN-Router} gap ($p=0.0151$).

\texttt{Short-Context} lowers Pass@1 to 44.0\%, supporting a longer reflection context; \texttt{Flash-Reflect} saves \$0.28 cost (\$4.64 versus \$4.92) but has lower observed Pass@1 (50.3\% versus 52.8\%), so we retain Pro reflection as the default. Additionally, \texttt{Max-Pass} and \texttt{Min-Cost} achieve similar Pass@1 and cost to the default, and neither differs significantly in Pass@1. 
% \emph{Thus, none of the tested alternatives improves on \compas{}'s default settings, validating the hyperparameter choices used throughout.}

\emph{\textbf{Takeaway.} None of the tested alternatives improves on \compas{}'s
default settings, validating the hyperparameter choices used throughout.}

% \section{Threats to Validity}
% \label{sec:threats}

% \textbf{Scope.} We test two benchmarks, three model families, and five seeds, but the model pool is not exhaustive and primary routing
% uses benchmark-provided difficulty labels; however, \texttt{XGBoost-Router}
% tracks this routing closely in a one-run sensitivity test.

% \textbf{Model selection.} When the cheap probe ties two pool models, its
% cost-aware rule selects the cheaper one; this caused the Devstral loss and may
% limit transfer to a new model family.

% \textbf{Search budget.} 1M budget outperforms in our setup, but it may require recalibration on other benchmarks or setups.

% \textbf{Implementation.} Although baseline reimplementations may differ from
% their original systems, we follow their published methods and evaluate every
% method with the same harness and documented token-cost accounting.

\section{Related Work}

\textbf{Search-Based Software Engineering $\times$ AI.}
\compas{} treats model, prompt, and decoding optimization as a search-based
software engineering (SBSE) problem \citep{harman2001sbse}: it uses
execution feedback to search for a configuration rather than hand-design
one. Within this line, \smac{} \citep{lindauer2022smac3} and
\promisetune{} \citep{chen2026promisetune} fit one global configuration,
whereas \alphaevolve{} \citep{novikov2025alphaevolve}, \shinkaevolve{}
\citep{lange2026shinkaevolve}, and \artemis{} \citep{brookes2025artemis}
evolve target programs. \gepa{} \citep{agrawal2025gepa} and \citet{madaan2023selfrefine} evolve prompts, while
\texttt{RoutingGen} \citep{li26intention} routes between fixed prompts by
difficulty. In contrast, \compas{} keeps the program and harness fixed,
jointly searches prompt and decoding settings, and retains a quality-cost
front for each difficulty group.

\textbf{LLM Configuration Optimization.}
LLM inference optimization also usually focuses on one dimension.
\routellm{} \citep{ong2025routellm}, \frugalgpt{} \citep{chen2023frugalgpt},
\hybridllm{} \citep{ding2024hybridllm}, \texttt{CSCR}
\citep{shirkavand2025costaware}, and \texttt{Router-R1}
\citep{feng2025routerr1} select models, whereas
\citet{arora2024hyperparams,du2025temperature,garces2025decoding,li2025temperature}
and \ecotune{} \citep{xu2025ecotune} tune decoding. \ecooptigen{}
\citep{wang2023costeffective} searches model, prompt, and decoding from a
fixed prompt list, but returns one configuration for every task. In contrast,
\compas{} combines model probing with reflection-based prompt-decoding search
and selects from a separate quality-cost front for each difficulty group.

\textbf{Agent Configuration and Harness Optimization.}
\gaivgc{} \citep{ga4gc2025} jointly searches agent hyperparameters and
prompt templates for one global configuration, whereas
\texttt{AgentConductor} \citep{wang2026agentconductor} builds a
difficulty-aware agent topology per task.
\openhands{} \citep{wang2025openhands}, \sweagent{} \citep{yang2024sweagent},
and harness evolution \citep{lin2026ahe} improve the tools, control logic,
or workflow. These approaches are complementary: \compas{} fixes the harness
and optimizes model, prompt, and decoding settings inside it, so the two can be used together
rather than treated as competing alternatives.

\section{Conclusion}

\compas{} shows that under a fixed harness, jointly selecting the model,
prompt, and decoding settings by task group improves the observed
quality-cost trade-off over routers, global tuners, and prompt-only optimizers
across splits, seeds, benchmarks, and model families. Its main
limitation is that each experiment only tested \compas{} on two same-family LLMs; future work could extend to cross-family search.

\bibliography{references}

% If the track requires an in-paper reproducibility checklist, uncomment:
% \input{ReproducibilityChecklist.tex}

\end{document}